\documentclass[onecolumn]{aa}
\usepackage{epsfig}

\usepackage{txfonts}
\begin{document}
   \title{Temperature fluctuations in \ion{H}{ii} regions:  $t^2$ for the
   two-phase model}
 
  \author{Y. Zhang
          \inst{1,2}
          \and
          B. Ercolano
             \inst{3}
          \and
          X.-W. Liu
           \inst{1}
          }

   \offprints{Y. Zhang}

   \institute{Department of Astronomy, Peking University,
               Beijing 100871, China\\
          \email{zhangy96@hkucc.hku.hk}
         \and
            Department of Physics, University of Hong Kong, Pokfulam Road, Hong Kong, China
         \and
    Harvard-Smithsonian Centre for Astrophysics, 60 Garden Street, Cambridge, MA 02138, USA\\
             }

   \date{Received; accepted}

   \abstract
{}
{We investigate temperature fluctuations in \ion{H}{ii} regions in terms of a
two-phase model, which assumes that the nebular gas consists of a hot and a
cold phase.  }
{We derive general formulae for $T$([\ion{O}{iii}]), the  [\ion{O}{iii}]
forbidden line temperature, and  $T$(\ion{H}{i}), the hydrogen Balmer jump
temperature, in terms of the temperatures of the hot and cold phases, $T_{\rm h}$
and $T_{\rm c}$.  }
{For large temperature differences, the values of $t^2$ required to account for the
observed difference between $T$([\ion{O}{iii}])  and $T$(\ion{H}{i}) are much
lower than those deduced using the classical formulae that assume random and
small amplitude temperature fluctuations.  One should therefore be cautious
when using a two-phase model to account for empirically derived $t^2$ values.
We present a correction of a recent work by Giammanco \& Beckman, who use a
two-phase model to estimate the ionization rate of \ion{H}{ii} regions by cosmic
rays.  We show that a very small amount of cold gas is sufficient to account
for $t^2$ values typically inferred for H~{\sc ii} regions.}
   {}



   \maketitle
%

\section{Introduction}

Temperature fluctuations in \ion{H}{ii} regions are a much-discussed problem.
Peimbert (\cite{p67}) investigated for the first time the effects of such
fluctuations on temperatures empirically derived from spectroscopic
observations and found that they may lead to higher electron temperatures being
derived from the collisionally excited [\ion{O}{iii}] nebular-to-auroral
forbidden-line ratio, $T$([\ion{O}{iii}]), than those derived from the Balmer
jump of the \ion{H}{i} recombination spectrum, $T$(\ion{H}{i}). If significant
temperature fluctuations exist in \ion{H}{ii} regions and yet are ignored in
the analysis, they may lead to underestimating ionic abundances calculated from
collisionally excited lines (CELs) (e.g. Esteban et al.  \cite{e02}).  The
parameter $t^2$ (Peimbert 1967) was introduced to quantitatively characterize
temperature fluctuations. Since then the parameter has been extensively used in
nebular studies. Under the conditions that $[T(r)-T_0]^2/T_0^2\ll1$, where
$T(r)$ is the local electron temperature and $T_0$ the average value weighted by
the square of density. The value of $t^2$ can be determined either by comparing
$T$([\ion{O}{iii}]) and $T$(\ion{H}{i}) or by comparing ionic abundances
derived from CELs and from recombination lines (RLs) (see Peimbert et al.
\cite{p04}, for further details).

Two-phase models, which approximate a nebula by two components of different
physical conditions, represent an over-simplified, yet frequently used method
of studying nebular physics (e.g. Viegas \& Clegg \cite{v94}; Zhang et al.
\cite{z05}). Using a two-phase model, Stasi{\' n}ska (\cite{s02} ) points out
that the classical picture of temperature fluctuations may be misleading under
certain conditions, and three parameters are needed to characterize temperature
inhomogeneities.  One of the open questions in the study of \ion{H}{ii} regions
is that values of $t^2$ derived from observations are consistently higher than
those predicted by photoionization models (Stasi{\' n}ska \cite{s00}).
Recently, Giammanco \& Beckman (\cite{g05}; GB05 thereafter) constructed a
two-temperature-phase model capable of explaining $t^2$ values deduced for a
number of \ion{H}{ii} regions by means of incorporating a component of cool
ionic gas ionized by cosmic rays.  However, $t^2$ values deduced from
observations cannot be applied directly to two-phase models. This is because
empirical values of $t^2$ deduced from observations were calculated from
formulae derived by assuming random and small-amplitude 
temperature fluctuations,
assumptions that are apparently broken for a two-phase model.

The purpose of the current work is to quantitatively study the relationship
between the $t^2$ values predicted by two-phase models and those measured by
observations. We show that, in the case of a very cold ionic gas component
embedded in a `normal' \ion{H}{ii} region, $t^2$ deduced from observations
using the empirical method may have significantly overestimated the real
values.


\section{Analysis}

According to Peimbert (\cite{p67}), for a given ionic species of number
density $N_i$, the thermal structure of an \ion{H}{ii} region can be
characterized by an average temperature $T_0$ and a mean square temperature
fluctuation parameter $t^2$, defined as
\begin{equation}
T_0=\frac{\int T_{\rm e}N_{\rm e}N_idV}{\int N_{\rm e}N_idV}
\end{equation}
and
\begin{equation}
t^2=\frac{\int (T_{\rm e}-T_0)^2N_{\rm e}N_idV}{T_0^2\int N_{\rm e}N_idV} \,,
\end{equation}
respectively.
Assuming $t^2\ll1$, $T_0$ and $t^2$ can be determined from measured 
$T$([\ion{O}{iii}]) and $T$(\ion{H}{i}) using relations,
\begin{equation}\label{p1}
T([\ion{O}{iii}])=T_0\left[1+\frac{1}{2}\left(\frac{9.13\times10^4}{T_0}-3\right)t^2\right]
\end{equation}
and
\begin{equation}\label{p2}
T(\ion{H}{i})=T_0(1-1.67t^2) 
\end{equation}
(Peimbert \cite{p67}; Garnett \cite{g92}). Esteban et al. (\cite{e02})
report $t^2$ values for a sample of \ion{H}{ii} regions. Their values are
reproduced in the first row of Table~\ref{est}.  Although they were obtained by
comparing ionic abundances derived from CELs and from RLs, we assume that they
are the same as would be deduced from from $T([\ion{O}{iii}])$ and
$T(\ion{H}{i})$ using Eqs.\,(3) and (4). The assumption is supported by studies
of Torres-Peimbert \& Peimbert (\cite{t03}), who find general agreement between
$t^2$ values inferred from the differences between $T([\ion{O}{iii}])$ and
$T(\ion{H}{i})$ and those inferred from the apparent discrepancies between CEL
and ORL abundances. Esteban et al. (\cite{e02}) did not provide values of
$T_0$. For the purpose of comparison, we follow GB05 and assume that
$T_0=T$([\ion{O}{iii}]), as given in the second row of Table~\ref{est}.
However, as pointed out previously by Stasi{\' n}ska (\cite{s02}), the
empirical method of estimating $t^2$ using Eqs.\,(3) and (4) may be invalid
under certain conditions. In the following, we show that in the two-phase
model, values of $t^2$ that are required to account for the measured
differences between $T([\ion{O}{iii}])$ and $T(\ion{H}{i})$ (or differences
between ORL and CEL abundances) may be much lower than that derived from the
empirical method.

In the framework of the two-phase model, which assumes that the electron
temperature structure of an \ion{H}{ii} region consists of a hot and a
cold phase, we follow GB05 and assume equal densities of the two phases and an
ionization fraction of unity for the hot gas. Electron temperatures are
designated as $T_{\rm h}$ and $T_{\rm c}$ for the hot and cold phases,
respectively. The intensity of an [\ion{O}{iii}] forbidden line transition of
wavelength $\lambda$ is given by
\begin{equation} \label{eq5}
I([\ion{O}{iii}])_\lambda \sim \int N({\rm O}^{2+})N_{\rm e}T_{\rm e}^{-1/2}\exp(-\Delta E/kT_{\rm e})dV \,,
\end{equation}
where $\Delta E$ is the excitation energy of the upper level. It follows that
\begin{eqnarray}
\nonumber
\frac{I([\ion{O}{iii}])_{4959,5007}}{I([\ion{O}{iii}])_{4363}}&\equiv&
C\times\exp\left\{\frac{33000}{T([\ion{O}{iii}])}\right\}\\
\nonumber
&=&C\times\frac{N({\rm O}^{2+})^{\rm h}N_{\rm e}^{\rm h}V^{\rm h}T_{\rm h}^{-1/2}\exp(-29200/T_{\rm h})+N({\rm O}^{2+})^{\rm c}N_{\rm e}^{\rm c}V^{\rm c}
T_{\rm c}^{-1/2}\exp(-29200/T_{\rm c})}{N({\rm O}^{2+})^{\rm h}N_{\rm e}^{\rm h}V^{\rm h}T_{\rm h}^{-1/2}\exp(-62200/T_{\rm h})+N({\rm O}^{2+})^{\rm c}N_{\rm e}^{\rm c}V^{\rm c}
T_{\rm c}^{-1/2}\exp(-62200/T_{\rm c})} \,, \\
\end{eqnarray}
where $C$ is a constant depending only on atomic data, and super- or subscript
`h' and `c' refer to quantities of the hot and cold phases, respectively. In
the cold phase, the ionization fraction of oxygen is expected to be lower than
that of hydrogen. As a reasonable approximation, we assume that $N({\rm
O}^{2+})^{\rm c}/N({\rm O})^{\rm c}=0.1 N({\rm H}^{+})^{\rm c}/N({\rm H})^{\rm
c}$. Our analysis is insensitive to this assumption since the cold phase
essentially contributes no [\ion{O}{iii}] forbidden line fluxes given its very
low temperature and the fact that emissivities of forbidden lines decline
exponentially with decreasing temperature (c.f Eq.~\ref{eq5}). We thus obtain
\begin{equation} \label{t1}
T([\ion{O}{iii}])=33000\ln^{-1}\left[\frac{T_{\rm h}^{-1/2}\exp(-29200/T_{\rm h})+0.1\theta x^2
T_{\rm c}^{-1/2}\exp(-29200/T_{\rm c})}{T_{\rm h}^{-1/2}\exp(-62200/T_{\rm h})+0.1\theta x^2
T_{\rm c}^{-1/2}\exp(-62200/T_{\rm c})}\right] \,,
\end{equation}
where, using the notation of GB05, $\theta$ is the mass ratio of the cold to
hot gas, and $x$ is the ionization fraction of hydrogen in the cold gas. It can be
easily seen that for $T_{\rm h}\gg T_{\rm c}$, $T([\ion{O}{iii}]) \sim T_{\rm
h}$. 

Similarly, from the flux of the Balmer jump,
\begin{equation}
I({\rm Bal}, 3646) \sim \int N({\rm H}^{+})N_{\rm e}T_{\rm e}^{-3/2}dV \,,
\end{equation}
 and the flux of H$\beta$,
\begin{equation}
I({\rm H}\beta) \sim \int N({\rm H}^{+})N_{\rm e}T_{\rm e}^{-5/6}dV \,,
\end{equation}

we have
\begin{equation}\label{t2}
T(\ion{H}{i})=\left(\frac{T_{\rm h}^{-3/2}+\theta x^2T_{\rm c}^{-3/2}}{T_{\rm h}^{-5/6}+\theta x^2T_{\rm c}^{-5/6}}\right)^{-3/2} \,,
\end{equation}
or
\begin{equation}\label{t22}
\theta x^2=\left(\frac{T_{\rm h}}{T_{\rm c}}\right)^{-3/2}\frac{T(\ion{H}{i})^{-2/3}T_{\rm h}^{2/3}-1}{1-T(\ion{H}{i})^{-2/3}T_{\rm c}^{2/3}} \,.
\end{equation}

 Equation (\ref{t2}) shows that $T(\ion{H}{i})$ weights towards $T_{\rm
c}$.  Following GB05, for a two-phase model, we obtain
\begin{equation}\label{t3}
T_0=\frac{T_{\rm h}+\theta x^2T_{\rm c}}{1+\theta x^2} \,,
\end{equation}
\begin{equation}\label{t4}
t^2=\theta x^2\left(\frac{T_{\rm h}-T_{\rm c}}{T_{\rm h}+\theta x^2T_{\rm c}}\right)^2 \,,
\end{equation}
\begin{equation}\label{v1}
\theta x^2=\frac{(T_{\rm h}-T_0)^2}{(T_0-T_{\rm c})^2}=\frac{T_0^2t^2}{(T_0-T_{\rm c})^2}  \,,
\end{equation}
and
\begin{equation}\label{v2}
T_{\rm h}=T_0+\frac{T_0^2t^2}{T_0-T_{\rm c}} \,.
\end{equation}
Therefore, for the two-phase scenario, $T_0$ and $t^2$ should be determined
from Eqs.~(\ref{t1}), (\ref{t2}), (\ref{t3}), and (\ref{t4}) instead
of from Eqs.~(\ref{p1}) and (\ref{p2}). The latter are only valid for random
and small amplitude fluctuations.

\begin{figure*}
\center
\epsfig{file=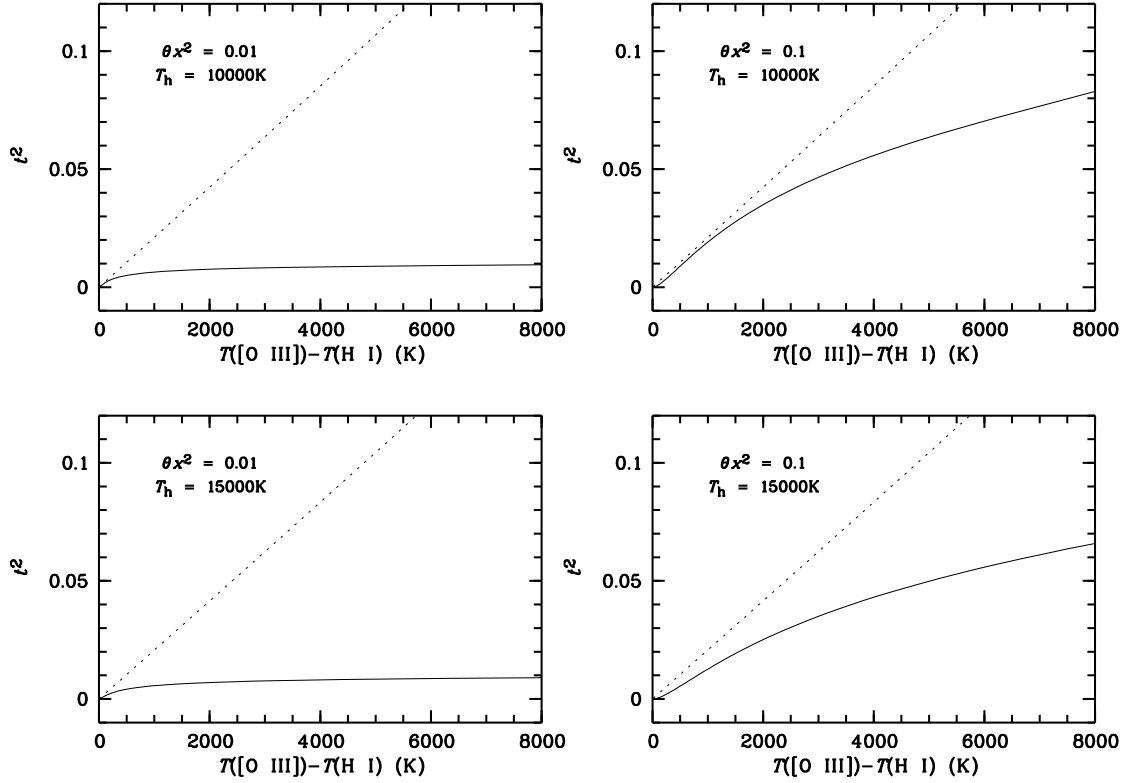,
height=10.5cm, bbllx=35, bblly=236, bburx=558, bbury=600, clip=, angle=0}
\caption{$t^2$ versus [$T($[\ion{O}{iii}]$)-T($\ion{H}{i}$)$], deduced from
the empirical method [Eqs.~(\ref{p1}) and (\ref{p2}), dotted lines] and for
two-phase model [Eqs.~(\ref{t1}), (\ref{t2}), (\ref{t3}), and (\ref{t4}),
solid lines]. {\it Left panels}: $\theta x^2$~=~0.01; {\it Right panels}: 
$\theta x^2$~=~0.1; {\it Upper panels}: $T_{\rm h}$~=~10000\,K; {\it Lower 
panels}: $T_{\rm h}$~=~15000\,K.}
\label{tsq}
\end{figure*}

 In Fig.~\ref{tsq} for given values of $\theta x^2$ and $T_{\rm h}$, we
compare $t^2$ as a function of $T([\ion{O}{iii}])-T(\ion{H}{i})$ derived from
the empirical method using Eqs.~(\ref{p1}) and (\ref{p2}), and that 
derived in the scenario of two-phase model using Eqs.~ (\ref{t1}), (\ref{t2}),
 and (\ref{t4}). The
plots show that, depending on $\theta x^2$ above a critical value of
$T([\ion{O}{iii}])-T(\ion{H}{i})$, the empirical method significantly
overestimates $t^2$, particularly for the case of small $\theta x^2$. The
amount of deviation is insensitive to the value adopted for $T_{\rm h}$.  In
addition, we find that as the temperature difference between the two phases is
larger than a critical value (typically $\sim6000$\,K), empirical $t^2$
deduced from observations can no longer be used at their face values to
constrain two-phase models, a point overlooked by GB05 as discussed in the
following section.

\section{New estimates of $\theta x^2$ for GB05 model}

GB05 showed that the ionization of cold neutral gas by cosmic rays may
significantly contribute to temperature fluctuations.  They used a two-phase
model to explain $t^2$ values obtained by Esteban et al.  (\cite{e02}) for a
number of H~{\sc ii} regions. However, the high temperature difference between
the two phases in GB05 model (see their Table~1) suggests that $t^2$ values
obtained by Esteban et al. cannot be applied directly to two-phase models.
Values of $\theta x^2$ derived by GB05 need to be re-considered.

We re-estimate $\theta x^2$ values for the sample of \ion{H}{ii} regions of
Esteban et al. (\cite{e02}). Following GB05, three values of $T_{\rm c}$ are
assumed, 100, 1000, and 4000\,K. Under these conditions, temperature in the 
cold gas is too low to collisionally excite the [\ion{O}{iii}] lines, and 
consequently Eq.~(\ref{t1}) can be simplified to
\begin{equation} \label{t5}
T([\ion{O}{iii}])=T_{\rm h} \,.
\end{equation}

  Substituting values of $T_0$ and $t^2$ given by Esteban et al.
(\cite{e02}) [hereafter referred as $T_{0,{\rm E}}$ and $t^2_{\rm E}$, in order
to distinguish them from those deduced from Eqs.~(\ref{t3}) and (\ref{t4})] into
Eqs.~(\ref{p1}) and (\ref{p2}), we obtain  $T$([\ion{O}{iii}])($t^2_{\rm
E}$,$T_{0,{\rm E}}$) and $T$(\ion{H}{i})($t^2_{\rm E}$,$T_{0,{\rm E}}$), which
are tabulated in Rows\,3 and 4 of Table~\ref{est}. Note that here we take the
same assumption by GB05 that $T_{0,{\rm E}}=T$([\ion{O}{iii}]). Then  $\theta
x^2$ can be determined from Eq.~(\ref{t22}), where
$T(\ion{H}{i})=T(\ion{H}{i})(t^2_{\rm E},T_{0,{\rm E}})$, $T_{\rm
h}=T([\ion{O}{iii}])(t^2_{\rm E},T_{0,{\rm E}})$.  Although the values of
$T$([\ion{O}{iii}])($t^2_{\rm E}$,$T_{0,{\rm E}}$) thus obtained are slightly
higher than the actual values deduced from observations (see Table~\ref{est}),
the resulting $\theta x^2$ are hardly affected.

In Table~\ref{est} we compare our $\theta x^2$ values to those of GB05 (given
in parentheses); for low values of $\theta x^2$ differences of up to a factor
of a hundred are found. It can easily be seen that the discrepancies increase
with decreasing temperature of the cold gas, as suggested by Fig.~\ref{tsq}.
For $T_{\rm c}=100$\,K, our derived values of $\theta x^2$ are very low,
suggesting that the values of $t^2$ reported by Esteban et al.  (\cite{e02})
can be explained by the existence of a very small amount of cold gas. 

Table~\ref{est} also gives $t^2$ and $T_0$ values derived from the
Eqs.~(\ref{t3}) and (\ref{t4}). As the Table shows, the real $t^2$ values are 
lower when $T_{\rm c} \ll T_0$ than those derived from the empirical method.
As a result, values of the cosmic ray ionization rate, $\zeta$, derived by GB05
have been grossly overestimated [c.f.  their Eq.~(17)]. Our conclusion is
consistent with the range of values inferred for the Orion nebula from
Gamma ray observations.

\begin{table*}
\centering \noindent \parbox{15cm}{ \centering \caption{Estimated values of
$t^2$, $T_0$, and $\theta x^2$ for $T_{\rm c}=100$, 1000, and 4000\,K for a
sample of \ion{H}{ii} regions. $T_{0,{\rm E}}$ and $t^2_{\rm E}$ are taken from
Esteban et al. (\cite{e02}). The numbers in parentheses are values 
derived by GB05 and included here for comparison.\label{est}}
\begin{tabular}{lcccc}
\hline\hline
{Object}        &  NGC~604  & NGC~5461 &  NGC~5471 &  NGC~2363 \\
\hline
$t^2_{\rm E}$   &  0.027 & 0.041 & 0.074 &  0.128 \\
$T_{0,{\rm E}}$ &  8150 &     8600 &  14100& 15700 \\
$T$([\ion{O}{iii}])($t^2_{\rm E}$,$T_{0,{\rm E}}$) & 9052 & 9943 &15913 & 18528\\
$T$(\ion{H}{i})($t^2_{\rm E}$,$T_{0,{\rm E}}$) & 7783  & 8011 & 12358 & 12344\\
$(t^2)_{100}$   &  0.00013  &  0.00016  &  0.00009  & 0.00013\\
$(t^2)_{1000}$  &  0.00413  &  0.00053  &  0.00312  & 0.00429\\
$(t^2)_{4000}$  &  0.02511  &  0.03503  &  0.02401  & 0.03540\\
$(T_0)_{100}$   &  9051     &  9941     &  15911    & 18526  \\
$(T_0)_{1000}$  &  9011     &  9884     &  15860    & 18445 \\
$(T_0)_{4000}$  &  8518     &  9370     &  15413    & 17719 \\
$(\theta x^2)_{100}$ & 0.00013(0.028) & 0.00017(0.042) &0.00010(0.075) &0.00013(0.130)\\
$(\theta x^2)_{1000}$& 0.00522(0.035)  &0.00659(0.052) &0.00356(0.086) &0.00480(0.146) \\
$(\theta x^2)_{4000}$& 0.08690(0.10)   &0.10664(0.14)  &0.04378(0.14)  &0.05905(0.23) \\
\hline\end{tabular}
}
\end{table*}

\section{Conclusion}

We have studied the relationship between values of $t^2$ predicted by a
two-phase model and those derived empirically from observations (empirical
method).  Our results show that the existence of extremely cold gas within
\ion{H}{ii} regions may lead to overestimated $t^2$ calculated from empirically
determined $T([\ion{O}{iii}])$ and $T(\ion{H}{i})$.  We stress that care should
be taken when using the two-phase model to study large temperature fluctuations
of \ion{H}{ii} regions. In this model, CELs are hardly produced by the cold
gas, which on the other hand makes a large contribution to the flux at the Balmer
jump, due to the $T_{\rm e}^{-3/2}$ dependence of $I({\rm Bal}, 3646)$.
Accordingly, the existence of a very small amount of cold material may lead to
a large discrepancy between $T([\ion{O}{iii}])$ and $T(\ion{H}{i})$. In other
words, in spite of its small mass, the existence of extremely cold material can
reproduce apparently large $t^2$ (as derived from the empirical method), much
larger than the actual value [as defined by Eq.~(2)].

Finally, we revisited the GB05 study of cosmic ray ionization as a mechanism
for creating temperature fluctuations in H~{\sc ii} regions. While this
provides a potential mechanism for creating cold ionized plasma in H~{\sc ii}
regions, we show that the values of $\zeta$ required to produce the ionization have
been overestimated in their treatment, due to the $t^2$ discrepancy discussed
above. Based on the formulae presented here, we re-estimated their model
parameters. The corrections are apparent, particularly in cases where
temperature of the cold gas component is low, resulting in lower values of $\zeta$
that agree better with the estimates for the Orion nebula published
in the literature.

\begin{acknowledgements} We thank the referee, Dr. C. Morisset, for helpful
comments that improved clarity of the paper. We would also like to thank
Dr.~Morisset for computing new values of $t^2$ and $T_0$, based on our
estimates of $\theta x^2$. Those values are now tabulated in Table~1.
YZ and XWL acknowledge support by NSFC grant \#10325312.

\end{acknowledgements}


\begin{thebibliography}{}
\bibitem[2002]{e02} Esteban, C., Peimbert, M., Torres-Peimbert, S.,
     \& Rodr{\' i}guez, M. 2002, \apj, 581, 241
\bibitem[1992]{g92} Garnett, D. R. 1992, \aj, 103, 1330
\bibitem[2005]{g05} Giammanco, C., \& Beckman, J. E. 2005, \aap, 437,
L11 (GB05)
\bibitem[1967]{p67} Peimbert, M. 1967, \apj, 150, 825
\bibitem[2004]{p04} Peimbert, M. , Peimbert, A., Ruiz, M. T., \&
Esteban, C. 2004, \apj, 150, 431
\bibitem[2000]{s00} Stasi{\' n}ska, G. 2000, RevMaxAA, 9, 158
\bibitem[2002]{s02} Stasi{\' n}ska, G. 2002, RevMaxAA, 12, 62
\bibitem[2003]{t03}
Torres-Peimbert, S., \& Peimbert, M. 2003, in IAU Symp. 209, Planetary Nebulae: Their Evolution and Role in the Universe, ed. P. R. Wood \& M. Dopita (San Francisco: ASP), p363
\bibitem[1994]{v94} Viegas, S. M., \& Clegg, R. E. S. 1994, \mnras, 271, 993
\bibitem[2005]{z05} Zhang, Y., Liu, X.-W., Liu, Y., \& Rubin, R. H. 2005, MNRAS,
358, 457
\end{thebibliography}
\end{document}